\documentclass[12pt]{amsart}

\usepackage{epsfig}

\begin{document}

\title[]{\bf Spherical Formulation for Diagramatic Evaluations
	     on a Manifold with Boundary}

\author[]{George Tsoupros \\
       {\em School of Physics}\\
       {\em The University of New South Wales}\\
       {\em NSW 2052}\\
       {\em Australia}
}
%\footnote{AMS Classification Numbers 49Q99, 81T15, 81T18, 81T20}
%\footnote{present e-mail address: georget@maths.usyd.edu.au}}
%\email{georget@maths.usyd.edu.au}
\thanks{present e-mail address: gts@phys.unsw.edu.au}

\begin{abstract}

The mathematical formalism necessary for the diagramatic evaluation of quantum 
corrections to a conformally invariant field theory for a self-interacting scalar 
field on a curved manifold with boundary is considered. The evaluation of quantum 
corrections to the effective action past one-loop necessitates diagramatic techniques. 
Diagramatic evaluations and higher loop-order renormalisation can be best 
accomplished on a Riemannian manifold of constant curvature accommodating a boundary 
of constant extrinsic curvature. In such a context the stated evaluations can be 
accomplished through a consistent interpretation of the Feynman rules within the 
spherical formulation of the theory for which the method of images allows. To this 
effect, the mathematical consequences of such an interpretation are analyzed and the 
spherical formulation of the Feynman rules on the bounded manifold is, as a result, 
developed.

\end{abstract}

\maketitle
%\newpage

{\bf I. Introduction}\\

The investigation of the effects generated on the dynamical behaviour of 
quantised matter fields by the presence of a boundary in the background 
geometry is an issue of central importance in Euclidean Quantum Gravity. This 
issue arises naturally in the context of any evaluation of radiative contributions to 
a semi-classical tunneling geometry and has been studied at one-loop level through 
use of heat kernel and functional techniques \cite{Barvinsky}, \cite{Esposito}. These 
methods were subsequently extended in the presence of matter couplings \cite{Moss}, 
\cite{Od}, \cite{Odin}, a context which, itself, allows for an improved effective 
action past one-loop order through use of renormalisation group techniques \cite{Odintsov}. 
Despite their success, however, such techniques have limited significance past one-loop 
order. Not only are explicit calculations of higher order radiative effects far more 
reliable for the qualitative assessment of the theory's dynamical behaviour under conformal 
rescalings of the metric but they are, in addition, explicitly indicative of boundary related 
effects on that behaviour. Yet, despite their significance no attempt has hitherto been made 
for the evaluation of higher loop-order contributions to the relevant effective action and, 
for that matter, to the semi-classical approximation relevant to the functional integral for 
quantum gravity. Such calculations necessarily rely on diagramatic techniques on a manifold 
with boundary. 

Fundamental in the calculational context relevant to such diagramatic techniques is 
the evaluation of the contribution which the boundary of the manifold has to the 
relativistic propagator of the relevant quantised matter field coupled to the manifold's 
semi-classical background geometry. A direct approach to such an evaluation is 
intractable. Specifically, in the physically important case of massless propagation of a 
scalar field conformally coupled to the background geometry of $ C_n$, that is to that of a 
n-dimensional manifold of constant curvature bounded by a $ (n-1)$-dimensional sphere - a 
case of direct relevance to quantum cosmology - any evaluation of the Green function 
associated with the bounded Laplace operator on the n-dimensional cap $ C_n$ involves 
expansions in terms of an incomplete set of spherical harmonics of fractional degrees. The 
latter constitute the eigenfunctions of that operator in contradistinction to the complete 
set of spherical harmonics of integer degrees on the n-dimensional sphere $ S_n$. Not only 
are such expansions intractable but, in addition, the concommitant emergence of fractional 
values for the degrees $ N$, which are physically associated with the angular momenta 
flowing through the relevant propagator, tends to obscure the physical interpretation of 
any perturbative calculation \cite{George}.

In view of such complications related to a diagramatic evaluation of higher order radiative 
contributions to the effective action a recent technique \cite{George} invoked the method of 
images in order to attain a reduction of the eigenvalue problem of the bounded Laplace 
operator on $ C_n$ to that of the unbounded Laplace operator on the covering manifold of 
$ C_n$, thus relating propagation on $ C_n$ to propagation on the entire $ S_n$. Such a 
manipulation allows for the exploitation of the established spherical formulation 
\cite{Drummond}, \cite{I.Drummond}, \cite{DrummondShore}, \cite{Shore}, \cite{G.Shore}
of the relevant quantum field theory. For that matter, a proper use of the 
method of images results perturbatively in quantum corrections to the semi-classical action for a 
conformal scalar field on $ C_4$ by exploiting the mathematical consequences of massless 
propagation in de Sitter space. As a preamble to a renormalisation program on $ C_4$ the 
stated technique was invoked in order to show that the vanishing result which dimensional 
renormalisation yields for the massless "tadpole" diagram in maximally symmetric spaces is 
no longer sustained in spaces of smaller symmetry, a situation which was shown to have the 
potential for the generation of conformally non-invariant counterterms at higher orders. 
Moreover, it was shown that the method of images is in absolute conformity with the physically 
intuitive expectation to the effect that no independent renormalisation is necessary. The volume 
related terms in the classical action receive infinite contributions from vacuum effects 
simultaneously with the boundary related terms. In this respect, the present technique revealed, 
in addition, a potential for non-trivial radiative generation of surface counterterms in the 
gravitational action \cite{George}.

The present work advances the stated technique in the direction of evaluating radiative
contributions to the effective action on a manifold with boundary. The advantage of working 
on a manifold of constant curvature stems from its underlying symmetry which allows for the 
exploitation of the method of images. To that effect, a diagramatic evaluation in the 
direction of perturbative renormalisation on $ C_4$ necessitates a consistent interpretation 
of the Feynman rules in the context of the mathematical framework set by the method of images. 
Consequent upon such an interpretation is a spherical formulation of the Feynman rules on 
$ C_4$. In contrast with the spherical formulation of the Feynman rules on the Euclidean $ S_4$ 
[8] the spherical formulation on $ C_4$ is substantially more intricate as a result of the 
smaller symmetry group and the presence of the boundary. Regardless of mathematical 
complications, however, the conceptual basis of the spherical formulation on $ C_4$ is always 
the massless scalar propagation on $ S_4$. As was advanced in [7] this premise becomes possible 
through the proper association of both the fundamental and boundary part of the relevant 
propagator on $ C_4$ with the propagator on $ S_4$. In what follows the mathematical consequences 
which this premise entails for diagramatic calculations on $ C_4$ will be 
explored within the context of dimensional regularisation \cite{'t Hooft}. Specifically, it will be 
shown that, although each divergence arising from the coincidence of space-time points receives 
contributions from the boundary, in conformity with intuitive expectation the pole structure of 
each diagram stems exclusively from the fundamental part of the propagator. Boundary-related 
contributions to divergences are, themselves, effected in configuration space by the volume
integral which, according to the Feynman rules, stems from vertex integration. Since vertex 
integration is dictated by the Feynman rules the stated volume integral has a general significance. 
All diagramatic calculations in configuration space necessitate an exact result for 
it. The integral over the volume of $ C_4$, however, no longer reduces to the orthonormality 
condition which emerges on $ S_4$. In pursuit of a spherical formulation for diagramatic evaluations 
on $ C_4$ the ensuing analysis will accomplish a reduction of the volume-related vertex integration 
to boundary terms which lend themselves to a consistent physical interpretation.

{\bf II. The method of images and Feynman diagrams on the bounded manifold}\\

As a matter of conveniece diagramatic calculations in relativistic field theory are usually
performed in momentum space. However, transform space calculations relevant to $ S_n$ exploit 
a complete set of spherical harmonics $ Y_{\alpha}^N({\eta})$ in $ n+1$ dimensions and are 
rather more involved \cite{Drummond}. Following the same calculational approach as that on 
$ S_n$, for that matter, any diagramatic evaluation on $ C_n$ will be advanced in 
configuration space as well. The spherical n-cap $ \it{C_n}$ itself will be considered as 
a manifold of constant curvature embedded in a $ (n+1)$-dimensional Euclidean space and 
bounded by a $ (n-1)$-sphere of constant positive extrinsic curvature $ K$ 
(diverging normals).
 
The propagator in configuration space for a conformal scalar field $ \Phi$ on a n-dimensional 
spherical cap $ C_n$ with the Dirichlet condition $ \Phi=0$ on the $ (n-1)$-dimensional
boundary $ \partial C_n$ has been attained through use of the method of images \cite{George}. 
With $ a_{\eta'}$ and $ a_B$ being the geodesic distances of space-time point $ \eta'$ and of 
$ \partial C_n$ from the cap's pole respectively the propagator was shown to be       
\footnote{with the symmetry factor relevant to the Haddamard function explicitly featured.}

\begin{equation}
D_{c}^{(n)}(\eta,{\eta}') = \frac{\Gamma(\frac{n}{2}-1)}{4\pi^{\frac{n}{2}}}\frac{1}{|{\eta}-{\eta}'|^{n-2}} 
- \frac{\Gamma(\frac{n}{2}-1)}{4\pi^{\frac{n}{2}}}\frac{1}{|\frac{a_{{\eta}'}}{a_B}{\eta}-
\frac{a_B}{a_{{\eta}'}}{\eta}'|^{n-2}}
\end{equation}
 
In this form the contribution which the propagator on $ S_n$ receives due to the presence 
of the boundary $ \partial C_n$ is explicit. The propagator on $ S_n$ is the fundamental part 
$ |{\eta}-{\eta}'|^{2-n}$ of the Green function which $ D_{c}^{(n)}(\eta,{\eta}')$ signifies 
on $ C_n$ \cite{George} and admits the expansion

\begin{equation} 
|{\eta}-{\eta}'|^{2-n} = \sum_{N=0}^{\infty}\sum_{\alpha=0}^{N}
\frac{1}{\lambda_N}Y_{\alpha}^N({\eta})Y_{\alpha}^N({\eta}')
\end{equation}
in terms of the complete set of spherical harmonics in $ n+1$ dimensions. The latter are 
eigenfunctions 

\begin{equation}
MY_{\alpha}^N(\eta) = \lambda_N Y_{\alpha}^N(\eta)
\end{equation}
of the Laplace operator on $ S_n$ 

\begin{equation}
M = D^2 - \frac{n(n-2)}{4a^2};~~~
D_a = (\delta_{ab} -
\frac{{\eta_a}\eta_b}{a^2})\frac{\partial}{\partial{\eta_a}}
\end{equation}
with eigenvalues $ \lambda_N$ expressed in terms of the degree $ N$, the dimensionality 
$ n$ and the radius $ a$ of the embedded Euclidean sphere as \cite{Drummond} 

\begin{equation}
\lambda_N = -\frac{(N+\frac{n}{2}-1)(N+\frac{n}{2})}{a^2}
\end{equation}

The boundary related contribution is, apparently, the second term in (1). In view of the 
analysis in \cite{George} this term, itself signifying propagation on $ S_n$ between the 
image points of $ \eta$ and $ \eta'$, admits the expansion  

\begin{equation}
|\frac{a_{{\eta}'}}{a_B}{\eta}-
\frac{a_B}{a_{{\eta}'}}{\eta}'|^{2-n} =
\sum_{N=0}^{N_0}\sum_{\alpha=0}^{N}
\frac{1}{\lambda_N}Y_{\alpha}^N(\frac{a_{{\eta}'}}{a_B}{\eta})
Y_{\alpha}^N(\frac{a_B}{a_{{\eta}'}}{\eta}')
\end{equation}
with the condition of truncation at $ N=N_0$ imposed by the non-vanishing geodesic 
separation $ |\frac{a_{{\eta}'}}{a_B}{\eta}'- \frac{a_B}{a_{{\eta}'}}{\eta}'|$ on 
$ S_n$ at the coincidence limit $ \eta \rightarrow \eta'$ on $ \it{C_n}$. 

In configuration space any diagramatic calculation on $ C_n$ involves powers of the 
propagator $ D_{c}^{(n)}(\eta,{\eta}')$ and, for that matter, products featuring powers of 
its fundamental part $ |{\eta}-{\eta}'|^{2-n}$ and of its boundary part 
$ |\frac{a_{{\eta}'}}{a_B}{\eta} - \frac{a_B}{a_{{\eta}'}}{\eta}'|^{2-n}$. The former are 
evaluated through use of the fundamental formula \cite{Drummond}

\begin{equation}
[(\eta - \eta')^2]^{\nu} = \sum_{N=0}^{\infty}\sum_{\alpha=0}^{N}
\frac{(2a)^{2\nu+n}{\pi}^{\frac{n}{2}}\Gamma(\nu+\frac{n}{2})\Gamma(N-\nu)}{
\Gamma(N+n+\nu)\Gamma(-\nu)}Y_{\alpha}^N({\eta})Y_{\alpha}^N({\eta}')
\end{equation}         
A rigorous mathematical derivation of (7) exploits the Funck-Hecke theorem \cite{Bateman} 
and is quite involved. However, this expression can also be seen to be a 
consequence of the eigenfunction expansion of $ |{\eta}-{\eta}'|^{2-n}$ in (2). The latter is, 
of course, a direct consequence of the fact that this expression is the configuration space 
propagator on $ S_n$ for the conformal scalar field. However, the expression 
$ |\frac{a_{{\eta}'}}{a_B}{\eta}- \frac{a_B}{a_{{\eta}'}}{\eta}'|^{2-n} $ which has been stated to 
admit the corresponding expansion in (6) is not formally a propagator on $ S_n$ 
even though in the context of the method of images it signifies propagation at finite geodesic 
separations on $ S_n$. This feature can be readily understood through the absence of the 
coincidence limit for which the propagator, being a Green function, should naturally allow. The 
only choice of the space-time points $ \eta$ and $ \eta'$ on $ C_n$ which can bring the image 
points $ \frac{a_{{\eta}'}}{a_B}{\eta}$ and $ \frac{a_B}{a_{{\eta}'}}{\eta}'$ into coincidence 
on $ S_n$ lies in the set of points which constitute the boundary $ \partial C_n$. 

As an immediate physical consequence of such a situation the expression 
$ |\frac{a_{{\eta}'}}{a_B}{\eta}- \frac{a_B}{a_{{\eta}'}}{\eta}'|^{2-n} $
remains always finite - as was mathematically expected from the boundary part of the 
Green function $ D_{c}^{(n)}(\eta,{\eta}')$. Even though it amounts, for that matter, to a 
finite multiplicative factor in any diagramatic evaluation it is worth noting that
its corresponding expansion in (6), although consistent in its upper bound of $ N_0$, has 
itself only approximate significance in terms of the spherical harmonics which it features. 
This is a consequence of the fact that the sets of spherical harmonics 
$ {Y_{\alpha}^N(\frac{a_{{\eta}'}}{a_B}{\eta})}$ and 
$ {Y_{\alpha}^N(\frac{a_B}{a_{{\eta}'}}{\eta}')}$ are not actually eigenfunctions of 
the unbounded Laplace operator on $ S_n$, cited in (4), except at the limit of 
$ a_{\eta'} \rightarrow a_B$ in which case (2) is recovered. In view of the 
approximate significance of the "eigenfunction expansion" which the boundary part of 
$ D_{c}^{(n)}(\eta,{\eta}')$ admits, an expansion corresponding to that in (7) 
for that term is also expected on condition of the same approximate significance      
   
\begin{equation}
[|\frac{a_{{\eta}'}}{a_B}{\eta}- \frac{a_B}{a_{{\eta}'}}{\eta}'|^2]^{\nu} = 
\sum_{N=0}^{N_0}\sum_{\alpha=0}^{N}
\frac{(2a)^{2\nu+n}{\pi}^{\frac{n}{2}}\Gamma(\nu+\frac{n}{2}+
\frac{1}{N_0})\Gamma(N-\nu+ \frac{1}{N_0})}{
\Gamma(N+n+\nu+ \frac{1}{N_0})\Gamma(-\nu)}Y_{\alpha}^N({\eta})Y_{\alpha}^N({\eta}')
\end{equation}  
where use has been made of \cite{George}

\begin{equation}
\sum_{\alpha=0}^{N}Y_{\alpha}^N(\frac{a_{{\eta}'}}{a_B}{\eta})
Y_{\alpha}^N(\frac{a_B}{a_{{\eta}'}}{\eta}') = \sum_{\alpha=0}^{N}
Y_{\alpha}^N({\eta})Y_{\alpha}^N({\eta}') 
\end{equation}
The inverse $ \frac{1}{N_0}$ of the upper limit related to the cut-off angular
momentum for image propagation in the arguments of the $ \Gamma$ functions ensures
the finite character of (8) at $ n \rightarrow 4$ and is expected on the grounds that 
the situation of scalar propagation towards the boundary - that is, that situation 
characterised by the limit $ a_{\eta'} \rightarrow a_B$ - results in 
$ N_0 \rightarrow \infty$ at the limit of vanishing geodesic separations 
$ \eta \rightarrow \eta'$ in which case, as discussed, the approximate expansion in 
(8) is reduced to the exact expansion in (7).         

Although a rigorous derivation of (8) is possible through a resolution of the spherical 
harmonics in terms of Gegenbauer polynomials and use of the Funck-Hecke theorem
such derivation is unnecessarily complicated for the purposes of diagramatic 
calculations in view of the fact that the final result amounts always to a finite multiplicative 
factor at the coincidence limit $ \eta \rightarrow \eta' $. As will become evident in 
the context of explicit diagramatic calculations the heuristic approach to transform space 
expansions of the boundary term outlined above is, for that matter, sufficient for the evaluation 
of the theory's dynamical behaviour at the high energy limit of vanishing geodesic 
separations on $ C_n$. In effect, in the context of dimensional renormalisation 
\cite{Drummond}, all possible divergences at the dimensional limit 
$ n \rightarrow 4$ stem exclusively from the fundamental-part related expansion 
in (7).

The preceding analysis reveals the mathematical origin and physical character of the pole
structures in configuration space. These structures, however, arise within the context of 
volume integrations which are a direct consequence of the Feynman rules. Specifically,  
any diagramatic calculation in configuration space entails integrations of the diagram's 
vertices over the relevant manifold's volume. 
In the present case powers of 
$ D_{c}^{(n)}(\eta,{\eta}')$ are integrated with respect to the space-time points 
$ \eta$ and $ \eta'$ over $ C_n$. Such integrations are expected to be ultimately 
responsible for the contributions which any divergence generated by the coincidence of 
space-time points receives from the boundary. Boundary-related divergences are expected
as the cummulative effect of reflection off the boundary of signals signifying propagation on
$ S_4$. Such divergences have been studied on a general manifold with boundary at one-loop
level through heat-kernel techniques \cite{AvraEspo}. Their presence in this theory will be explicitly confirmed 
past one-loop level in the context of future diagramatic calculations. Since a condition for such
calculations as well as the objective of  the present analysis is the spherical formulation of the 
Feynman rules on $ C_4$ it is imperative that the stated volume integrations be evaluated to 
an exact expression. To that effect it can readily be seen from (7) and (8) that such integrations 
invariably eventuate in the expression $ \int_Cd^{n}{\eta}Y_{\alpha}^N(\eta)Y_{\alpha'}^{N'}(\eta)$. 
Even though this integral features a product of spherical harmonics on $ S_n$ its evaluation necessitates 
caution in view of the fact that the range of integration is strictly over $ C_n$ as a consequence of 
which the usual orthonormality condition 

\begin{equation}
\int_S d^{n}{\eta}Y_{\alpha}^N(\eta)Y_{\alpha'}^{N'}(\eta) =
\delta_{NN'}\delta_{\alpha \alpha'}
\end{equation}
is precluded. It would be desirable, for that matter, to attempt a reduction of this
integral to one on the $(n-1)-$dimensional spherical hypersurface of constant positive 
extrinsic curvature which constitutes the boundary $ \partial C_n$. The boundary 
hypersurface corresponds to a specific value $ \theta_n^0$ of the angle relevant to the 
$ n-$th component of the space-time vector $ \eta$. This calculational direction is 
suggested by the resolution of the spherical harmonics in $ n+1$ dimensions in terms 
of Gegenbauer polynomials $ C_m^{\nu}(cos{\theta})$ of degree $ m~ \epsilon~ Z$ and 
order $ \nu;~~ 2\nu~ \epsilon~ Z$ \cite{Bateman} 

\begin{equation}   
Y_{\alpha}^N(\eta) = a^{-\frac{n}{2}}e^{\pm im_{n-1}\theta_1}\prod_{k=0}^{n-2}
(sin{\theta_{n-k}})^{m_{k+1}}
C_{m_k-m_{k+1}}^{m_{k+1}+\frac{1}{2}(n-k)}(cos{\theta_{n-k}})
\end{equation}
which can be seen to indicate an orthonormality condition on 
$ \partial C_n \equiv S^{\theta_n^0}$. The derivation of that condition will be realised 
in due course. It is, nevertheless, important at this point to stress that the use of 
spherical harmonics in this project as well as in \cite{George} is identical to that made 
in \cite{Drummond} which differs from that of \cite{Bateman} in that the latter is 
dimensionless in length dimensions. Consistency with (10), for that matter, dictates the 
multiplicative factor of $ a^{-\frac{n}{2}}$ which (11) features. It is worth noting, in addition, 
that the degree of homogeneity $ N$ of the spherical harmonics in $ n+1$ dimensions is, 
as expected, the sum-total of the degrees $ m_k-m_{k+1}$ of the individual Gegenbauer 
polynomials which enter (11). For that matter, $ N=m_0$. 

In this context, the integral in question can, instead, be evaluated through the eigenvalue 
equation (3). For that matter, it is

$$
\int_C d^{n}{\eta}Y_{\alpha}^N(\eta)Y_{\alpha'}^{N'}(\eta) = $$

\begin{equation}
\frac{1}{\lambda_{N'} - 
\lambda_{N}}[\int_C d^{n}{\eta}Y_{\alpha}^N(\eta)MY_{\alpha'}^{N'}(\eta) - 
\int_C d^{n}{\eta}Y_{\alpha'}^{N'}(\eta)MY_{\alpha}^{N}(\eta)] 
\end{equation}
However, (4) implies 

$$
\int_C d^{n}{\eta}Y_{\alpha}^N(\eta)MY_{\alpha'}^{N'}(\eta) - 
\int_C d^{n}{\eta}Y_{\alpha'}^{N'}(\eta)MY_{\alpha}^{N}(\eta) = $$

\begin{equation}
\int_C d^{n}{\eta}Y_{\alpha}^N(\eta)D^2Y_{\alpha'}^{N'}(\eta) - 
\frac{n(n-2)}{4a^2}\int_C d^{n}{\eta}Y_{\alpha}^N(\eta)Y_{\alpha'}^{N'}(\eta)
- [(N,{\alpha}) \leftrightarrow (N',{\alpha'})]
\end{equation}
the left-hand side of which, immediately, reduces to 

%\begin{equation}
%\int_C d^{n}{\eta}Y_{\alpha}^N(\eta)MY_{\alpha'}^{N'}(\eta) - 
%\int_C d^{n}{\eta}Y_{\alpha'}^{N'}(\eta)MY_{\alpha}^{N}(\eta) = 
$$
\int_C d^{n}{\eta} D_a[Y_{\alpha}^N(\eta)D_aY_{\alpha'}^{N'}(\eta)] - 
\int_C d^{n}{\eta} D_a[Y_{\alpha'}^{N'}(\eta)D_aY_{\alpha}^{N}(\eta)] $$
%\end{equation}

This expression is now directly amenable to the desired reduction on 
$ S^{\theta_n^0}$. A direct application of the divergence theorem on $ C_n$, 
for that matter, further reduces the left-hand side of (13) to 

%$$
%\int_C d^{n}{\eta}Y_{\alpha}^N(\eta)MY_{\alpha'}^{N'}(\eta) - 
%\int_C d^{n}{\eta}Y_{\alpha'}^{N'}(\eta)MY_{\alpha}^{N}(\eta) = $$

$$ 
\oint_{\partial C} d^{n-1}{\eta}n_pY_{\alpha}^N(\eta)D_pY_{\alpha'}^{N'}(\eta) -
\oint_{\partial C} d^{n-1}{\eta}n_pY_{\alpha'}^{N'}(\eta)D_pY_{\alpha}^{N}(\eta) 
$$
whereupon, with $ n_p$ being the unit vector normal to $ S^{\theta_n^0}$, yet 
another application of the divergence theorem yields

$$
\int_C d^{n}{\eta}Y_{\alpha}^N(\eta)MY_{\alpha'}^{N'}(\eta) - 
\int_C d^{n}{\eta}Y_{\alpha'}^{N'}(\eta)MY_{\alpha}^{N}(\eta) = $$

\begin{equation}
\oint_{\partial C} d^{n-1}{\eta}[-KY_{\alpha}^N(\eta)Y_{\alpha'}^{N'}(\eta) - 
2n_pY_{\alpha'}^{N'}(\eta)D_pY_{\alpha}^{N}(\eta)]
\end{equation}
where use has been made of $ \partial \partial C_n = 0$ and the fact that the extrinsic 
curvature tensor $ K_{ij}$ of $ \partial C_n \equiv S^{\theta_n^0}$ whose trace $ K > 0$ 
appears manifestly in (14) is $ K_{ij}=\frac{1}{2}(D_{i}n_{j}+D_{j}n_{i})$. Together (14) and 
(5) finally reduce (12) to 

\begin{equation}
\int_C d^{n}{\eta}Y_{\alpha}^N(\eta)Y_{\alpha'}^{N'}(\eta) = Aa^2
\oint_{\partial C} d^{n-1}{\eta}[KY_{\alpha}^N(\eta)Y_{\alpha'}^{N'}(\eta) + 
2n_pY_{\alpha'}^{N'}(\eta)D_pY_{\alpha}^{N}(\eta)]
\end{equation}
with 

\begin{equation}
A = \frac{1}{(N'+\frac{n}{2}-1)(N'+\frac{n}{2}) - (N+\frac{n}{2}-1)(N+\frac{n}{2})};~~N \neq N' 
\end{equation}
Although still incomplete this reduction of the volume integral on the boundary 
allows for an assessment of certain qualitative aspects of the theory. In addition to the 
radius $ a$ of the embedded $ \it{C_n}$ it features, as expected, the extrinsic curvature of 
$ \partial C_n$. Evidently, it is exactly this feature of any diagramatic calculation in 
configuration space which, as will be explicitly shown in the context of perturbative 
calculations, ensures the announced simultaneous renormalisation of boundary and surface 
terms in the effective action at any specific loop order. Moreover, the supplemented condition 
$ N \neq N'$ stems from the fact that any coincidence between $ N$ and $ N'$ enforces, through 
(13), a vanishing result on the boundary integral in (15). However, since the denominator in 
its multiplicative factor $ A$ also vanishes for $ N = N'$ the volume integral in (15) remains 
always finite. It should, already, be evident that this effect ensures the absence of infra-red 
divergences in $ C_4$, a result which will also be confirmed in the context of explicit 
calculations. 

In order to complete the reduction of the volume integral in (15) on the boundary 
$ \partial C_n$ it is necessary that the two terms featured in the boundary integral be 
examined separately. Applying the resolution in (11) to the product of the two spherical 
harmonics in the boundary integral of (15) and taking into consideration that \cite{Bateman}  

\begin{equation}
\int_{C}{d^n}{\eta} = a^n
\int_0^{2\pi}d\theta_1\int_0^{\pi}d\theta_2sin\theta_2\int_0^{\pi}d\theta_3sin^2\theta_3...
\int_0^{\theta_n^0}d\theta_nsin^{n-1}\theta_n
\end{equation}

results in 

%\newpage

$$
\oint_{\partial C} d^{n-1}{\eta}Y_{\alpha}^N(\eta)Y_{\alpha'}^{N'}(\eta) = $$

\begin{equation}
a^{-1}(sin{\theta_n^0})^{m_1+l_1}C_{N-m_1}^{m_1+\frac{n-1}{2}}(cos{\theta_n^0})
C_{N'-l_1}^{l_1+\frac{n-1}{2}}(cos{\theta_n^0})
\oint_{\partial C} d^{n-1}{\eta}Y_{a}^{m_1}(\eta^{\theta_n^0})Y_{a'}^{l_1}(\eta^{\theta_n^0})  
\end{equation}      
with {$Y_{a}^{m}(\eta^{\theta_n^0})$} being the spherical harmonics of degree $ m$ 
defined strictly on the $ (n-1)$-dimensional spherical hypersurface $ S^{\theta_n^0}$ 
which constitutes $ \partial C_n$ and being, as can be deduced from (11), of length 
dimensionality equal to $ \frac{-n+1}{2}$. Moreover, the geometry of the embedded 
Euclidean sphere makes it evident that 

\begin{equation}
m = Ncos{\theta_n^0}
\end{equation}
This is the case because the degree $ N$ of $ Y_{\alpha}^N$ is, through (2) and (6), 
physically associated with the angular momentum flowing through the relevant propagator
\cite{George}. In the case of (19) the geodesic propagation associated with $ N$ occurs 
on that hypersurface which constitutes the boundary $ \partial C_n$. This might appear to 
be counterintuitive in view of the absence of any propagation on $ \partial C_n$. Such an
absence, however, characterises exclusively the propagator on $ C_n$ expressed in (1). 
By contrast, the method of images renders, through (2) and (6), the massless scalar 
propagation relevant to $ S_n$ \cite{George}. The stated absence on $ \partial C_n$ is
the exclusive result of the combination of those two expressions in the context of (1).   

In effect, a direct application of (10) in $ (n-1)$-dimensions reduces (18) to

$$
\oint_{\partial C} d^{n-1}{\eta}Y_{\alpha}^N(\eta)Y_{\alpha'}^{N'}(\eta) = $$

\begin{equation}
a^{-1}(sin{\theta_n^0})^{m_1+l_1}C_{N-m_1}^{m_1+\frac{n-1}{2}}(cos{\theta_n^0})
C_{N'-l_1}^{l_1+\frac{n-1}{2}}(cos{\theta_n^0})\delta^{m_1l_1}\delta_{aa'} 
\end{equation}
on the understanding that \cite{Bateman}

\begin{equation}
\delta_{\alpha \alpha'} = \delta^{m_1l_1}\delta_{m_2l_2}...\delta_{m_{n-1}l_{n-1}}
= \delta^{m_1l_1}\delta_{a a'}
\end{equation}    
Eq. (20) could have also been derived from (15), (11) and (17) through use 
of the orthonormality relation for the Gegenbauer polynomials \cite{Bateman} (properly 
normalised in conformity with \cite{Drummond})

\begin{equation}     
\int_0^{\pi}C_m^{p}(cos{\theta})C_l^{p}(cos{\theta})(sin{\theta})^{2p}d{\theta} = 
\delta_{lm}
\end{equation}
The result expressed by (20) features explicitly the expected orthonormality condition 
on $ S^{\theta_n^0}$ and is, for that matter, the final reduction on $ \partial C_n$ of 
the first surface integral in (15). It becomes evident, at this stage, that the reduction of  
the complete orthonormality condition on $ S_n$ expressed by (10) to the partial 
orthonormality condition on $ C_n$ expressed by (20) in the context of (15) is responsible 
for the contributions which any divergence on $ S_4$ may receive due to the presence of 
$ \partial C_4$. Although this effect will become transparent in light of explicit calculations
a heuristic argument for it can be presented on the following lines. As stated, all pole 
structures in the context of dimensional regularisation stem exclusively from the 
fundamental part of the propagator. However, any finite contribution to the effective action 
on $ S_4$ which emerges as a result of a cancellation of poles is unattainable on 
$ C_4$ if the pole in the denominator only stems from a Gamma function the argument of which features 
the degree $ N$

	$$ \frac{\Gamma(\epsilon)}{\Gamma(N+\epsilon)};~~ \epsilon \rightarrow 0~~ as~~ n \rightarrow 4 $$
The reason is that the orthonormality condition (10) on $ S_n$ enforces, in that case, 
the result $ N=0$ in the argument of the corresponding Gamma 
function in the denominator and, for that matter, the stated cancellation at 
$ n \rightarrow 4$ \cite{Drummond}. The same result is, nevertheless, unattainable in the 
context of (15) as a consequence of which the pole in the numerator persists on $ C_4$ 
for all $ N \neq 0 $ terms in the relevant expansion stemming from (7). Since the 
replacement of (10) by (15) responsible for that detraction is, itself, a direct consequence of the 
presence of the boundary $ \partial C_n$ it becomes evident that all boundary-related contributions 
to divergences already present on $ S_n$ are, necessarily, effected by the volume integral which stems 
from vertex integration in the configuration space of $ C_n$. 

The evaluation of the second surface integral is 
substantially more involved as the presence of the directional covariant derivative in 
it precludes any orthonormality condition. In order to proceed it is necessary to, 
explicitly, express that derivative in spherical coordinates. Considering that

\begin{equation}
(\frac{\partial}{\partial \eta_a}) = 
(\frac{\partial}{\partial a},~~\frac{1}{a}\frac{\partial}{\partial 
\theta_{1}},~~\frac{1}{asin{\theta_1}}\frac{\partial}{\partial \theta_2},...,
%\frac{1}{Rsin{\theta_1}...sin{\theta_{n-2}}}\frac{\partial}{\partial \theta_{n-1}},~~
\frac{1}{asin{\theta_1}...sin{\theta_{n-1}}}\frac{\partial}{\partial \theta_{n}})
\end{equation}     
and that for vectors constraint on the embedded n-sphere $ \eta^2 = a^2$ it is 

\begin{equation}
(\eta_a) = (a,~~0,...,0)
\end{equation} 
it follows through (4) that the directional derivative normal to the boundary amounts to

\begin{equation}
{n_pD_p}_{\mid \theta_n^0} = 
\frac{1}{asin{\theta_1}...sin{\theta_{n-1}}}\frac{\partial}{\partial \theta_{n}}_{\mid \theta_n^0}
\end{equation}             
The directional derivative is now expressed in spherical coordinates and can, as such, 
be applied in the context of (11). It is  

$$
{n_pD_p}{Y_{\alpha}^N(\eta)}_{\mid \theta_n^0} = $$
\begin{equation}
e^{\pm im_{n-1}\theta_1}\frac{a^{-\frac{n}{2}}}{asin{\theta_1}...sin{\theta_{n-1}}}
B\prod_{k=1}^{n-2}(sin{\theta_{n-k}})^{m_{k+1}}
C_{m_k-m_{k+1}}^{m_{k+1}+\frac{n-1}{2}-\frac{1}{2}k}(cos{\theta_{n-k}}) 
\end{equation}
with 

$$  
B = [\frac{m_{1}}{2}(sin{\theta_n^0})^{m_{1}-1}(2cos{\theta_n^0})
C_{N-m_{1}}^{m_1+\frac{n-1}{2}}(cos{\theta_n^0})~~- $$

\begin{equation}
(sin{\theta_n^0})^{m_{1}+1}(2m_1+n-1)
C_{N-m_{1}-1}^{m_1+\frac{n-1}{2}+1}(cos{\theta_n^0})]
\end{equation} 
where use has been made of \cite{Drummond} 

\begin{equation}
\frac{d^k}{dt^k}C_m^p(t) = \frac{2^k\Gamma(p+k)}{\Gamma(p)}C_{m-k}^{p+k}(t)
\end{equation}
Finally, combining (26) and (27) with (11) and integrating over $ C_n$ results in the 
desired reduction of the second surface integral in (15) 

$$    
\oint_{\partial C} d^{n-1}{\eta}[2n_pY_{\alpha'}^{N'}(\eta)D_pY_{\alpha}^{N}(\eta)] = 
a^{-n-1}\tilde{B}~~ 
\times
$$

$$
\oint_{\partial C} d^{n-1}{\eta}
\prod_{k=1}^{n-2}(sin{\theta_{n-k}})^{m_{k+2}+l_{k+2}-1}
C_{m_{k+1}-m_{k+2}}^{m_{k+2}+\frac{n-1}{2}-\frac{1}{2}k}(cos{\theta_{n-k}}) 
C_{l_{k+1}-l_{k+2}}^{l_{k+2}+\frac{n-1}{2}-\frac{1}{2}k}(cos{\theta_{n-k}})~~
\times $$

\begin{equation}
\frac{e^{i(\pm m_{n-1}\mp l_{n-1})\theta_1}}{sin{\theta_1}}
\end{equation}
with 

$$  
\tilde{B} = 2[m_1(sin{\theta_n^0})^{m_{1}+l_{1}-1}(cos{\theta_n^0})
C_{N-m_{1}}^{m_1+\frac{n-1}{2}}(cos{\theta_n^0})~~ - $$

\begin{equation}
(sin{\theta_n^0})^{m_{1}+l_{1}+1}(2m_1+n-1)C_{N-m_{1}-1}^{m_1+\frac{n-1}{2}+1}(cos{\theta_n^0})]
C_{N'-l_{1}}^{l_1+\frac{n-1}{2}}(cos{\theta_n^0})
\end{equation} 
This time, however, no orthonormality condition emerges for this term. Combining (29) with (17) reveals
that the surface integral in the former ``falls apart'' as

$$
[asin{\theta_n^0}]^{n-1}
\int_0^{\pi}C_{m_{1}-m_{2}}^{m_{2}+\frac{n-2}{2}}(cos{\theta_{n-1}})C_{l_{1}-l_{2}}^{l_{2}+
\frac{n-2}{2}}(cos{\theta_{n-1}})[sin{\theta_{n-1}}]^{m_{2}+l_{2}+n-3}d{\theta_{n-1}} \times ...$$

\begin{equation}
\int_0^{\pi}C_{m_{n-2}-m_{n-1}}^{m_{n-1}+\frac{1}{2}}(cos{\theta_{2}})C_{l_{n-2}-l_{n-1}}^{l_{n-1}+
\frac{1}{2}}(cos{\theta_{2}})[sin{\theta_{2}}]^{m_{n-1}+
l_{n-1}}d{\theta_{2}}\int_0^{2\pi}
\frac{e^{i(\pm m_{n-1}\mp l_{n-1})\theta_1}}{sin{\theta_1}}
\end{equation}
Clearly the presence of the directional derivative in this term has reduced by one the power in 
the weight relevant to the inner product in (22). Again, this is a consequence of the reduction 
of the incipient symmetry under the Euclidean de Sitter group $ SO(n+1)$ to the smaller symmetry 
relevant to $ C_n$. As a result of the preceding analysis then, the volume integral which is 
associated with the Feynman rules in the configuration space relevant to the conformal scalar 
field on $ C_n$ is expressed by (15), (20), (29) and (31) with the additional conditions in (16), 
(21) and (30).

{\bf III. Conclusions}\\

All diagramatic calculations in the configuration space associated with the dynamical behaviour 
of a conformal scalar field on a manifold with boundary can be best accomplished on a manifold 
of constant curvature with a boundary of constant extrinsic curvature. The advantage of working 
on a Euclidean n-dimensional spherical cap $ C_n$ stems from its underlying symmetry which 
allows for the exploitation of the method of images. The latter, in turn, allows for a considerable 
association of the dynamical behaviour of the quantum field on $ C_n$ to that of the same field on 
the Euclidean n-sphere $ S_n$.
%In effect, the concommitant spherical formulation renders each Green 
%function an exact function of the Ricci scalar $ R$. This result has both formal and perturbative 
%significance. 
In the context of the spherical formulation for which that association allows all configuration 
space-related diagramatic calculations on $ C_n$ have been shown to necessitate the result of the 
exact expansion in (7) as well as that of the expansion in (8). The approximate character of the 
latter renders it more reliable in situations of scalar propagation towards the boundary 
$ \partial C_n$. Regardless of the context of scalar propagation, however, the approximate 
significance of that mathematical manipulation has been shown to be inconsequential to the 
evaluation of the divergences at the high energy limit of vanishing geodesic separations as the 
image propagation, which (8) physically signifies on $ S_n$, is characterised by the absence of such 
vanishing separations between the associated image points. Moreover, the method of images allows for an 
exact reduction on the boundary $ \partial C_n$ of the, otherwise intractable, volume integral associated 
with the expansions in (7) and (8). The result relevant to all diagramatic calculations is     

\newpage

$$
\int_C d^{n}{\eta}Y_{\alpha}^N(\eta)Y_{\alpha'}^{N'}(\eta) = AKa
(sin{\theta_n^0})^{m_1+l_1}C_{N-m_1}^{m_1+\frac{n-1}{2}}(cos{\theta_n^0})
C_{N'-l_1}^{l_1+\frac{n-1}{2}}(cos{\theta_n^0})\delta^{m_1l_1}\delta_{aa'}~~ + $$

$$ 
(sin{\theta_n^0})^{n-1}A\tilde{B}
\prod_{k=1}^{n-2}\int_0^{\pi}C_{m_k-m_{k+1}}^{m_{k+1}+\frac{n-2}{2}}(cos{\theta_{n-k}})
C_{l_k-l_{k+1}}^{l_{k+1}+\frac{n-2}{2}}(cos{\theta_{n-k}})[sin{\theta_{n-k}}]^{m_{k+1}+l_{k+1}+n-3}
d{\theta_{n-k}}\times $$

$$
\int_0^{2\pi}\frac{e^{i(\pm m_{n-1}\mp l_{n-1})\theta_1}}{sin{\theta_1}}
$$
with

$$
m_1 = Ncos{\theta_n^0}~~;~~l_1 = N'cos{\theta_n^0}
$$
and confirms the theory's potential for simultaneous renormalisation of volume and surface terms
in the effective action at arbitrarily high loop-orders. This general result is, in addition, suggestive
of the potential contributions which divergences already present on $ S_4$ receive due to the presence of 
the boundary $ \partial C_n$.  

These results are the essential expression of the spherical formulation for the Feynman rules
on $ C_4$. As such they constitute the mathematical framework necessary for higher loop renormalisation on a 
manifold with boundary. In the sequel to this project an explicit calculation of the zero point function to 
order three in the loop expansion will demonstrate in this respect the merit of the technique hitherto 
developed.

\end{document}